\documentclass[aps]{revtex4}

\begin{document}

\title{Coarse-grained Description of 
Polymer Blends as Interacting Soft-Colloidal Particles}
\author{G. Yatsenko, E. J. Sambriski, M. G. Guenza}
\affiliation{Department of Chemistry,
Institute of Theoretical Science,
University of Oregon,
Eugene, OR 97403}


\begin{abstract}
We present a theoretical approach which maps polymer blends onto mixtures of 
soft-colloidal particles. The analytical mesoscale pair distribution
functions reproduce well data from united atom molecular dynamics simulations of
polyolefin mixtures without
fitting parameters. The theory exactly recovers the analytical expressions
for density and concentration
fluctuation structure factors of soft colloidal mixtures (liquid alloys).
\end{abstract}

\maketitle

\section{Introduction}
Polymer blends are systems of fundamental scientific interest. Their 
structural and dynamical properties change as a function of the
proximity to thermodynamic conditions of phase
separation (i.e., their spinodal curve). Blends have many 
practical applications since new materials can be produced
with specific physical and chemical properties by simply
mixing polymer melt components that have the desired characteristics.
Finally, polymeric materials used in daily life are usually mixtures of 
polymers with different chain
length and/or different local structure. Understanding how the mixing of polymer
melts modifies their properties has been a relevant and longstanding scientific and
technological goal in polymer physics and engineering.\cite{Balsara,book,Muller}
From a theoretical perspective, relevant
work in this direction has been developed by several groups 
in recent years.\cite{schweiz94,Schweiz97,Freed,Freed1,Freed2,Lipson}

A great deal of information on the correlation between local (intra- and intermolecular)
structure and global 
fluid properties has been obtained by computer simulations
of polymer mixtures.\cite{Muller,Binder,Mattice,Maranas,Grest,grest}
One of the challenges in simulating polymer blends is the large range 
of length- and timescales
that need to be investigated. Blend properties strongly depend on
the local bond lengthscale (short timescale), since mixtures of polymers with different local
chemical structure can lead to dramatically different physical properties, including, for
example,
an opposite trend in  demixing with temperature.
On the other end, properties need to be investigated on the large lengthscale (long timescale)
of concentration fluctuations, which diverges approaching
the spinodal decomposition. In this regime, the box size of the system simulated becomes
the upper limit in resolution.

To expedite computer simulations, it is useful 
to renormalize the system
by adopting a coarse-grained description of
the liquid. The goal is to have a formalism that allows us to simulate
macroscopic behavior while retaining information about the detailed atomistic chemical
structure of polymer chains. The overall behavior of blend materials has to be 
related, through a well-defined
and reversible procedure, to the local chemical structure of the system.
As a  first step, a convenient approach is to introduce a 
united atom  (UA) description of the polymer chain: a commonly adopted method
in simulations of polymer melts and blends.\cite{Maranas,Grest,Paul}
A more drastic renormalization at the lengthscale of the polymer size was proposed by 
Dautenhahn and Hall\cite{Hall} and later on by Murat and Kremer\cite{Kremer},
who
mapped a blend of
polymer chains onto a liquid of colloidal particles interacting through phenomenological
soft-core potentials.
Recently, Hansen and coworkers have developed a rigorous numerical 
methodology to derive
an effective potential for polymers in solution using liquid 
state theory.\cite{Hansen,Hansen1}

An analytical solution of the c.o.m.\ mean-force
potential for a melt was derived by one of the authors.\cite{marinamacrom,marinaprl} 
The potential is qualitatively in agreement with the physical behavior observed in 
simulation data. For example,
when compared with the classical solution by Flory and Krigbaum,\cite{Flory}
it has the advantage of correctly predicting greater
interpenetration between two
chains  with increasing
degree of polymerization, chain stiffness, and/or liquid density.
However, since 
the derivation does
not include explicitly chain connectivity, the analytical solution
reproduces the short-scale simulation behavior after fitting the 
value of the intermolecular
potential at complete interpolymer 
overlap.\cite{marinaprl}
More recently we derived an  analytical expression for the potential of mean-force
in polymer melts,
which reproduces well, and with no adjustable parameters, data from UA Molecular Dynamics (MD)
simulations of chains 
with different degree of polymerization
and local chemical structure.\cite{meltprl}
This coarse-grained potential
is an explicit function of 
atomistic chain parameters, such as  degree of polymerization and local semiflexibility,
thereby correctly bridging different lengthscales of interest.

It is clearly advantageous to have an analytical form of the potential versus
a numerical one, obtained form the inversion of computer
simulation data.  Since the mean-force potential is a free energy, it
depends on the thermodynamic state of the system considered. A numerical
calculation of the potential would require us to perform numerical
simulations for each thermodynamic state of our system, defeating the
purpose of adopting a coarse-grained model to reduce computational time.
In this spirit,  Krakoviack et al. recently derived an analytical potential 
for polymers in solution.\cite{Hansen2}

In this paper we extend the melt theory to derive an analytical  
expression 
for the effective
interaction potential acting between the 
centers of mass (\mbox{c.o.m.}) of two polymer chains in a blend.
We start from a first-principles liquid state description, thus limiting our theory
to the miscible region
of the phase diagram. However, it
includes the buildup of concentration fluctuations as the system
approaches its spinodal decomposition.
It is known that the detailed nature of single-phase
blend correlations (pair distribution functions)
is sensitive to system-specific factors such as liquid density, temperature, blend
composition, and
differences in effective unit size and local architectural details of the components.
Moreover, many blends of practical relevance are characterized by strong asymmetries in
local chemical semiflexibility and architecture. These
effects influence local entropic packing of molecules and can lead to phase
separation upon heating in a blend characterized by a lower critical solution temperature.
Our analytical expression describes the interaction betweeen
c.o.m.\ of a pair of polymers in a blend as a function of all the characteristic
parameters that govern blend structure and dynamics.

The derived mesoscale pair distribution functions effectively map the polymeric
liquid onto a fluid of interacting
soft-colloidal particles, thereby correctly
reproducing fluctuations
in number density and
concentration, i.e. structure factors, and recovering the well-known equations for the
compressibility and 
the dilation factor of colloidal mixtures.\cite{Bhatia,KirkwoodBuff} 
Moreover, the theory agrees well,
without fitting parameters,
with data from UA-MD computer simulations\cite{Grest,grest}
for polyolefin blends with
different local stiffness, architecture, and blend composition.

Good agreement between the theory of mesoscopic colloidal particle mixtures and
microscopic UA computer simulations of polymer blends
supports
the validity of our analytical renormalization procedure for
polymer mixtures.
The theory 
makes predictions 
on the evolution of the effective interpolymer potentials and chain packing
as a function of temperature and blend composition, as well as polymer local structure,
semiflexibility, and degree of polymerization.

This paper is divided in the following way: in Sec.\ II, we derive 
monomer pair distribution functions for a binary asymmetric polymer blend
where differences in bond length, degree of polymerization, and composition are
explicitly taken into account.  This is followed by a comparison with 
computer simulation data.  In Sec. III, we present a renormalized description of the
polymer fluid by deriving 
c.o.m.\ pair distribution functions.  A brief discussion
of the associated mean-force potential at chain-overlap is given in Sec.\ IV.  We then
show how our formalism can readily be cast into the language of soft-colloidal systems
in Sec.\ V.  In Sec.\ VI we summarize model calculations for our
renormalized description of a binary blend and make comparisons with simulation data.
A brief discussion concludes the paper.

\section{Monomer Pair Distribution Functions for Asymmetric Polymer Blends}
In this section we develop the theory for monomer pair distribution
functions in an asymmetric polymer blend, which will be inputs to our renormalization
approach. The theory is then tested against data from UA-MD simulations.
We include the effect of microscopic and macroscopic parameters through  
a generalized form of the monomer-monomer pair distribution function, as derived from
the thread limit in the Polymer Reference Interaction Site Model (PRISM) 
liquid-state theory 
by Curro, Schweizer, and coworkers.\cite{PRISM,PRISM1,PRISM2} 
The choice of the theory representing
the monomer pair distribution
function is arbitrary, and other analytical approaches 
could have been chosen as well.\cite{Freed,Lipson}
The original thread model is implemented here to include the dependence on 
asymmetries in
semiflexibility following a somewhat different procedure 
than in the original work.\cite{TangSch} No
closure approximations are adopted in this derivation, and the miscibility 
parameter $\chi$ enters as an input to the theory, defining the proximity of the
system to its demixing transition (the spinodal temperature). In principle, this approach
can describe systems with either Lower (LCST) or Upper (UCST) Critical Solution Temperature.

In the thread model, a single chain is represented as an infinite thread of vanishing  
thickness with hard
core monomer diameter
$d\rightarrow 0$ and segment number density $\rho\rightarrow \infty$, while $\rho d^3$
is finite and of order unity. Although the number of segments is infinite and their
size approaches zero, the overall polymer dimension, given by its radius-of-gyration $R_g$
is finite, and enables a direct comparison with data from computer simulations of 
finite-size polymer chains.
This simple chain model gives a very rough description of the liquid on the local
segment scale, since it completely averages out solvation shells in the pair distribution
function. However, this approximated description of the local structure
becomes correct in the limit of long polymer chains, where the fine structure of
the distribution function does not appear.

The thread model is the analog of the model investigated in field theory approaches.
Although analytically tractable, it
has proven to be quite successful in describing 
properties of polymer melts, block copolymers, and blends.\cite{PRISM,block,block1} 
The main reason for 
its success is that it 
correctly captures the onset of 
``correlation hole" effects\cite{DeGennes} in the monomer pair distribution functions at the 
lengthscale of the overall polymer size $R_g$.
The presence of the ``correlation hole" is
a trademark of 
polymer fluid structure and governs the physics of the system at the chain lengthscale.
Since in our mesoscopic description of polymer liquids $R_g$ is the lengthscale of interest,
the thread model appears to be adequate for the purpose of
the theory developed here.

We start from a binary blend of $A$ and $B$
homopolymers, having
degrees of polymerization $N_A$ and $N_B$, and
unit lengths $\sigma_A$ and $\sigma_B$, respectively. The polymer volume fraction is
$\phi=n_AN_A/(n_A N_A+n_B N_B)$, where $n_A$ is the number of molecules of type $A$ in the
mixture.
Non-bonded interactions are quantified by the single $\chi$
parameter 
which describes 
the ``monomer" interchange energy for the blend.
The generalized Ornstein-Zernike (OZ) matrix relation for total
site-site correlation functions of a  binary fluid mixture in Fourier 
space\cite{HansenMcd,schweiz89} is
\begin{eqnarray}
\mathbf{H}{\it(k)} = \mathbf{\Omega}{\it (k)} \mathbf{C}{\it (k)} \left[\mathbf{\Omega}{\it (k)} +
\mathbf{H}{\it (k)}\right] \ ,
\label{OZ}
\end{eqnarray}
where the matrices in Eq.(\ref{OZ}) are block matrices of rank
$2$. Here, $H_{\alpha\beta}(k)$, $C_{\alpha\beta}(k)$, and
$\mathit{\Omega}_{\alpha\beta}(k)$ are Fourier transforms of the
corresponding correlation functions;
for example, 
\begin{eqnarray}
H_{\alpha\beta}(k)=\frac{4\pi}{k} \int_0^\infty  
r H_{\alpha\beta}(r) \sin(kr) dr \ .
\end{eqnarray}
In real space,
$H_{\alpha\beta}(r)= \rho_{\alpha}\rho_{\beta}h_{\alpha\beta}(r)$
is the (chain-averaged) site-site total distribution function matrix;
$C_{\alpha\beta}(r)$ is the intermolecular direct correlation function
matrix; and
$\mathit{\Omega}_{\alpha\beta}(r) = \rho_{\alpha}\omega_{\alpha}(r)\delta_{\alpha\beta}$
is the intrachain structure factor matrix, 
with $\alpha,\beta \in \{A,B\}$ and $\rho_{\alpha}=n_\alpha N_\alpha /V$
the number density of monomers $\alpha$ inside volume $V$.
The site averaged intrachain correlations are described by the
Pad\'{e} approximant to the
Debye function (which introduces only a 15\% maximum error to the 
exact expression\cite{DoiEdw})
\begin{eqnarray}
\omega_{\alpha}(k)\approx\frac{N_{\alpha}}{1+k^2R_{g\alpha}^2/2} \ .
\label{intra} 
\end{eqnarray}
Although approximated, Eq.(\ref{intra}) enables
an analytically tractable solution to the intermolecular c.o.m.\-c.o.m.\
potential, which is the ultimate goal of the theory developed.
Moreover, the use of the Pad\'{e} approximant gives good agreement with
simulations for the total pair distribution function in both real and 
reciprocal spaces\cite{meltprl}.
Agreement between the Debye and the Pad\'{e} forms improves as $N$ increases.

Partial structure factors are derived from Eq.(\ref{OZ}) and 
$\mathbf{S}{\it (k)}=\mathbf{\Omega}{\it (k)}+ \mathbf{H}{\it (k)}$ as\cite{PRISM}
\begin{eqnarray}
&S_{AA}(k)&=\rho\phi\omega_A(k)[1-\rho(1-\phi)\omega_B(k)]/\Lambda(k)   
 \nonumber \\
&S_{BB}(k)&=\rho(1-\phi)\omega_B(k)[1-\rho\phi\omega_A(k)]/\Lambda(k) 
\label{stfBB1}\\
&S_{AB}(k)&=\rho^2\phi(1-\phi)\omega_A(k)\omega_B(k)C_{AB}(k)/\Lambda(k)
\label{stfAB1} \nonumber
\end{eqnarray}
with $\rho$ the total segment number density of the blend and
\begin{eqnarray}
\Lambda(k)=1&-&\rho\phi\omega_A(k)C_{AA}(k)-\rho(1-\phi)\omega_B(k)C_{BB}(k)+\nonumber\\
&+&\rho^2\phi(1-\phi)\omega_A(k)\omega_B(k)[C_{AA}(k)C_{BB}(k)-C_{AB}^2(k)] \ .
\label{lambda}
\end{eqnarray}
%
The direct correlation function $C_{\alpha \beta}(k)$ represents the fluid-averaged 
intermolecular pair potential,
which is short-ranged or independent of $k$ for $k\sigma \ll 1$.  
Since the lengthscale of interest for a mesoscopic description ($R_g$) 
is larger than $\sigma$, the direct correlation function  can be assumed to be
independent of $k$,
\begin{eqnarray}
C_{\alpha\beta}(k)\approx C_{\alpha\beta}(k=0)=C_0^{\alpha\beta}-
{\epsilon}_{\alpha\beta}/{\rho} \ .
\label{C0}
\end{eqnarray}
The first term in Eq.(\ref{C0}), $C_0^{\alpha\beta} < 0$, is the hard core 
repulsion of particles
at direct contact (the potential $(k_BT)^{-1} v(r) \approx - C(r)$), 
while the second term, ${\epsilon}_{\alpha\beta}/{\rho} < 0$, defines
the ``tail" interaction potential 
between species $\alpha$ and 
$\beta$. 
$C_0^{\alpha\beta}$ governs the pure athermal packing of monomers in 
the blend, so it contains the entropic contribution that can lead
to demixing in the high-temperature region. Consistently with 
the thread model 
we assume that the hard core components of 
the direct correlation function
become $C_0^{\alpha\beta}\approx C_0$ for 
any combination of $A$ and $B$. 
Introducing Eq.(\ref{intra}) and
Eq.(\ref{C0}) in Eqs.(\ref{stfBB1}) enables the factorization of partial
structure factors 
into two
separate contributions related to density and 
concentration fluctuations
\begin{eqnarray}
&S_{AA}(k)&\approx \frac{12\rho\phi^2}{\sigma_{AB}^2}
\left ( \frac{1-\phi}{\phi}\frac{1}{k^2+\xi_{\phi}^{-2}}+
\frac{\gamma^2}{k^2+\xi_{\rho}^{-2}} \right )
\label{stfAA3} \nonumber \\
&S_{BB}(k)&\approx \frac{12\rho(1-\phi)^2}{\sigma_{AB}^2}
\left(\frac{\phi}{1-\phi}\frac{1}{k^2+\xi_{\phi}^{-2}}+
\frac{\gamma^{-2}}{k^2+\xi_{\rho}^{-2}}\right)
\label{stfBB3}\\
&S_{AB}(k)&\approx \frac{12\rho\phi(1-\phi)}{\sigma_{AB}^2}
\left(-\frac{1}{k^2+\xi_{\phi}^{-2}}+\frac{1}{k^2+\xi_{\rho}^{-2}}\right) \ , \nonumber
\end{eqnarray}
where $\sigma_{AB}^2=\phi \sigma_B^2+(1-\phi) \sigma_A^2$.
Density fluctuations are characterized by the correlation length
\begin{eqnarray}
\xi_{\rho}^{-2} = -12 \rho C_0 
\left[\frac{\phi}{\sigma_A^2} + \frac{1-\phi
}{\sigma_B^2}\right] \ .
\label{xiroa}
\end{eqnarray}
Concentration fluctuations  
follow an incompressible-like relation $S_{AA}=S_{BB}=-S_{AB}$,\cite{DeGennes} and
are characterized by the correlation length
\begin{eqnarray}
\xi_{\phi}=\frac{\sigma_{AB}}{\sqrt{24\phi(1-\phi)(\chi_s-\chi)}}
= \frac{\xi_{cA}}{\sqrt{1-\chi/
\chi_s}}\frac{\sqrt{1-\phi+\phi\gamma^2}}{\sqrt{1-\phi+\phi
\mu}} \ ,
\label{xif2}
\end{eqnarray}
which diverges at the spinodal temperature. 
Here, $\chi/\rho$ is the analog of the Flory-Huggins interaction parameter
given by
$\chi=\epsilon_{AB}-(\epsilon_{AA}+\epsilon_{BB})/2$. 
At the spinodal temperature $\chi \rightarrow \chi_s$, where 
$\chi_s=[2 N_A \phi]^{-1}+ [2 N_B (1-\phi)]^{-1}$.
The
ratio between $A$ and $B$ chain lengths is given by $\mu = N_A/N_B$, while
$\gamma=\sigma_B/\sigma_A$ defines the ratio of local stiffness
between species $A$ and $B$. Since the definition of $\gamma$ is arbitrary,
for convenience we assume that the stiffest
component in the mixture is $B$. Also,
\begin{eqnarray}
\xi_{c\alpha}=R_{g\alpha}/\sqrt{2} \ ,
\label{xic}
\end{eqnarray}
is the correlation hole lengthscale for polymer $\alpha$. 
Eq.(\ref{xif2}) implies that 
asymmetry in the segment length between two components ($\gamma \neq 1$)
increases the range of concentration fluctuations.

The total site-site correlation function is derived from the partial structure
factors as
\begin{eqnarray}
\rho^2\phi^2h_{AA}^{mm}(k)&=&S_{AA}(k)-\rho\phi\omega_A(k) 
\label{OZAA} \nonumber \\
\rho^2(1-\phi)^2{h}_{BB}^{mm}(k)&=&S_{BB}(k)-\rho(1-\phi){\omega}_B(k)
\label{OZBB}\\
\rho^2\phi(1-\phi){h}_{AB}^{mm}(k)&=&S_{AB}(k) \ , \nonumber
\label{OZAB}
\end{eqnarray}
where we introduced the label $mm$ to distinguish monomer correlation
functions from those of the c.o.m., which are derived in Section III.
Eqs.(\ref{OZBB}) reduce in real space to 
the generalized thread expressions
for the monomer pair distribution functions of asymmetric polymer blends  
\begin{eqnarray}
h_{AA}^{mm}(r) 
&=& \frac{3}{{\pi\rho r \sigma_{AB}^2}} \left[ 
         \frac{{1-\phi}}{\phi} e^{-r/\xi_{\phi}}
         + \gamma^2 e^{-r/\xi_{\rho}}
-\frac{1}{{\phi}} \frac{\sigma_{AB}^2}{\sigma_A^2}
e^{-r/\xi_{cA}} \right] \ ,
\label{graaas} \nonumber \\
h_{BB}^{mm}(r) & = & \frac{3}{{\pi\rho r \sigma_{AB}^2}}
\left[\frac{\phi}{1-\phi}
e^{-r/\xi_{\phi}}
+\gamma^{-2}
e^{-r/\xi_{\rho}}
-\frac{1}{1-\phi} \frac{\sigma_{AB}^2}{\sigma_B^2}
e^{-r/\xi_{cB}}\right] \ ,
\label{grbbas} \\
h_{AB}^{mm}(r) & = & \frac{3}{{\pi\rho r \sigma_{AB}^2}}
\left[-e^{-r/\xi_{\phi}}+e^{-r/\xi_{\rho}}\right] \ .
\label{grabas} \nonumber
\end{eqnarray}
Eqs.(\ref{graaas}) describe the distribution of elementary
units in a binary blend as a function of the intramolecular
structure and thermodynamical parameters. 
In these expressions, the three lengthscales of interest $\xi_{\rho}$, $\xi_c$, and
$\xi_\phi$ appear uncoupled.
Even in the athermal regime where $\chi \ll \chi_s$ the mismatch in size ($\mu\neq 1$) 
and/or in 
local semiflexibility ($\gamma \neq 1$) between 
the two components can lead to non-uniform mixing, 
corresponding to
$h_{AA}(r)\neq h_{BB}(r) \neq h_{AB}(r)$.
However, if the system is totally symmetric ($N_A=N_B$ and $\sigma_A=\sigma_B$),
Eqs.(\ref{grbbas}) 
recover the PRISM equations for a totally symmetric blend, which for athermal conditions
recover the melt thread-model equation.\cite{PRISM}

The equations just derived 
cannot be compared 
directly with simulation data of monomeric pair distribution functions since
the thread model does not describe the presence of solvation shells in $g(r)=h(r)+1$
as observed in simulations. 
However, a good
estimate for the accuracy of the derived expressions can be achieved by
comparing with simulations
the number of particles included within a sphere of 
radius $r$, as
\begin{eqnarray}
n_{\beta}(r)-\delta_{\alpha\beta}=4\pi\rho_{\beta}\int^r_0 {r^2g_{\alpha\beta}
(r)dr}
\label{nofr}
\end{eqnarray}
with $\rho_{\beta} \in \{\phi\rho,(1-\phi)\rho\}$ for $\beta \in \{A,B\}$,
respectively.
Tests of our monomer pair distribution functions 
are data from UA-MD computer simulations.\cite{Grest,grest}
The simulation procedure has been described in
detail in recent papers\cite{Grest,grest} and will not be discussed here.
Systems investigated are blends of isotactic polypropylene (iPP), 
head-to-head polypropylene (hhPP), polyisobutylene (PIB), syndiotactic polypropylene (sPP),
and 
polyethylene (PE).
Simulations and blend parameters are described in the Table.

In the simplified theoretical framework of the previous
section, we enforced the limit of large polymer chains, which leads to a single value for the
density fluctuation lengthscale, independent of the blend component. 
Here we re-introduce finite-size effects, local semiflexibility, and
branching, which are specific to each component, through a melt-like description of the local
density fluctuation lengthscale as
$\xi_{\rho \alpha} ^{-1}=\pi\rho\sigma_{\alpha}^2/3+\xi_{c \alpha}^{-1}$
and $\alpha\in\{A,B\}$. 
For the cross pair distribution, we have $\xi_{\rho \alpha \beta}^{-1}=
\pi\rho \sigma_{AB}^2/3+\xi_{c \alpha \beta}^{-1}$.
The statistical segment length is calculated from the polymer radius of gyration as
$\sigma_A=R_{gA}\sqrt{6/N_A}$ and $\sigma_B=R_{gB}\sqrt{6/N_B}$. These formulas hold
in the limit of long flexible chains obeying Gaussian statistics. Systems investigated here
include linear and branched polyolefins with small, densely-packed pendant groups.
It is known that, in general, highly branched polyolefins are more flexible
than their linear counterparts having the same degree of
polymerization.\cite{Bates92,wein95} Moreover,
blends of polymers with closely-spaced and/or small-sized
pendant groups are characterized by less 
efficient local packing of units. These effects produce a locally-disordered
liquid,\cite{Maranas2004,Rajasek} consistent with 
the ``random packing" of a Gaussian chain and with the smooth shape of
the thread monomer distribution function adopted here.

Most blends considered in the simulations are far from their critical
temperature.
For some samples, an equation for the $\chi$ parameter is reported
in the literature, as summarized in the Table, and has been used here for
the calculation of the concentration fluctuation correlation length.
However, the size of effective sites considered in the
analysis of the experimental data is in general different for the two components,
$\sigma_A\neq \sigma_B$. In these cases we normalize the value of 
$\chi$  by the average
number of sites per monomer\cite{grest} so as to be consistent with the
Flory-Huggins spinodal equation adopted here.

To describe samples for which an equation of $\chi$ is unknown, we perform
calculations in the athermal limit 
($\chi \ll \chi_s$) where concentration
fluctuations are minimized. Most of the blends considered here have
cross contributions $AB$ which could be 
approximated fairly well by an arithmetic or a
geometric average of the self terms, $AA$ and $BB$. This is a
characteristic property of homogeneous liquids, and for blends it occurs in the athermal 
regime where the components are randomly mixed (if entropic packing effects
are not dominant).

Using the procedure just described, we calculated the
number of units included in a sphere of radius $r$ around a tagged polymer
$\alpha$.  
The theory shows good agreement
with simulation data already at an intermolecular distance of a few
site lengths (Fig.1). While the thread model cannot reproduce simulation data for 
$r \rightarrow 0$, it correctly describes the average liquid structure at 
intermediate and large ($R_g$) lengthscales, which are the lengthscales of
interest in our mesoscopic description. We observe that the specific PRISM 
thread model adopted here
tends to overestimate the number of self and 
cross contacts. For a UCST blend, the self-correlation functions $AA$ and $BB$
are best described by the equations in athermal conditions ($chi<<chi_s$),
while the cross-correlation contribution $AB$ is
best represented by the equation in
thermal conditions.
These trends are visible also in c.o.m.\ total pair distribution
functions presented
in Section VI, a feature that appears to depend on the choice of 
monomer pair distribution functions.

\section{Center-of-mass soft-core potential in polymer blends}
To derive an analytical expression for the c.o.m.\ intermolecular
distribution functions in a polymeric mixture, we extend to polymer blends a procedure 
outlined by Krakoviack, Hansen, and Louis\cite{Krakoviack} 
for homopolymer solutions. 
While the contribution due to real sites (monomers
or effective units)
is averaged in the usual PRISM-like fashion,
the c.o.m.\ is included in the OZ relation given by Eq.(\ref{OZ}) as
an effective ``auxiliary" site. 
For a binary fluid mixture, the matrices in Eq.(\ref{OZ})
become block matrices of rank $4$.
We define the site-site total distribution function matrix as
\begin{eqnarray}
\mathbf{H}{\it (k)} & = &
\left[\begin{array}{*{2}{c}}
\mathbf{H}{\it ^{mm}(k)} & \mathbf{H}{\it ^{mC}(k)} \\
\mathbf{H}{\it ^{Cm}(k)} & \mathbf{H}{\it ^{CC}(k)} 
\end{array}\right] \ .
\label{hmm1}
\end{eqnarray}
In real space $\mathit{H}_{\alpha\beta}^{mm}(r)= \rho_{\alpha}\rho_{\beta}
h_{\alpha\beta}^{mm}(r)$ is
the (chain-averaged) site-site total distribution function matrix;
$\mathit{H}_{\alpha\beta}^{mC}(r)= \rho_{\alpha}\rho_{ch,\beta}h^{mC}_{\alpha\beta}(r)$
and $\mathit{H}_{\alpha\beta}^{Cm}(r)= \rho_{ch,\alpha}
\rho_{\beta}h^{Cm}_{\alpha\beta}(r)$ are
the site-c.o.m.\ total distribution functions where $\rho_{ch,\alpha}$ is the number 
density of chains of species $\alpha$; finally,
$\mathit{H}_{\alpha\beta}^{CC}(r)= \rho_{ch,\alpha}\rho_{ch,\beta}h_{\alpha\beta}(r)$ is
the c.o.m.-c.o.m.\ total distribution function matrix.
The direct correlation function matrix,
\begin{eqnarray}
\mathbf{C}{\it (k)} & = &
\left[
\begin{array}{*{2}{c}}
\mathbf{C}{\it ^{mm}(k)} & 0 \\
0 & 0  
\end{array} \right] \ ,
\label{cmm1}
\end{eqnarray}
includes the condition that direct correlation functions between auxilary
and real sites as well as between two
auxiliary sites are negligible, while the ``monomer" 
direct correlation functions are defined in Eq.(\ref{C0}). 
The intramolecular distribution function matrix is given by
\begin{eqnarray}
\mathbf{\Omega}{\it (k)} & = & \left[ 
\begin{array}{*{2}{c}}
          \mathbf{\Omega}{\it ^{mm}(k)} & \mathbf{\Omega}{\it ^{mC}(k)} \\
          \mathbf{\Omega}{\it ^{Cm}(k)} & \rho_{ch} 
\end{array} \right] \ ,
\label{omega} 
\end{eqnarray}
with $\mathit{\Omega}^{mm}_{\alpha\beta}(r) = \rho_{\alpha}\omega_{\alpha}^{mm}(r)\delta_{\alpha\beta}$
the site-site intrachain structure factor matrix, and
$\mathit{\Omega}^{mC}_{\alpha\beta}(r) = \rho_{ch,\alpha}\omega_{\alpha}^{mC}(r)\delta_{\alpha\beta}$
the site-c.o.m.\ intrachain structure factor matrix.
The block in the
upper-left quadrant in each matrix defines correlations between ``monomer" units, whereas the
block in the lower-right quadrant defines correlations between
centers of mass.
The two remaining off-diagonal quadrants contain information on the
correlation between real and auxiliary sites.
By using the definition of the static structure factor $\mathbf{S}{\it (k)}= 
\mathbf{\Omega}{\it (k)}+\mathbf{H}{\it (k)}$,
the ``monomer" structure factor matrix
recovers
Eqs.(\ref{OZAA}) as it should. It also correctly
reproduces 
Eqs.(\ref{stfAB1}) when the relation 
$\mathbf{S}(k) = [\mathbf{1}-\mathbf{\Omega}(k) 
\mathbf{C}(k)]^{-1}\mathbf{\Omega}(k)$ is enforced. 
If the two species in the binary blend are assumed to be identical, Eqs.(\ref{hmm1}-\ref{omega}) 
recover
equations for a polymer melt.\cite{meltprl}
In general we have that $h_{\alpha \beta}^{Cm}(k) =h_{\beta \alpha}^{mC}(k)$
while $h_{\alpha \beta}^{Cm}(k) \neq h_{\alpha \beta}^{mC}(k) $ with $\alpha \neq \beta$.
Using these conditions, Eqs.(\ref{OZ}) are solved to obtain the c.o.m.\
total correlation functions in Fourier space 
\begin{eqnarray}
h_{\alpha \beta}(k) =  \left(\frac{\omega_{\alpha}^{mC}(k)\omega_{\beta}^{mC}(k)}
{\omega_{\alpha}^{mm}(k)\omega_{\beta}^{mm}(k)}\right)h_{\alpha \beta}^{mm}(k) \ .
\label{hcc}
\end{eqnarray}
Eq.(\ref{hcc}) formally connects c.o.m.\ distribution functions to the
``monomer" intra- and intermolecular distribution functions, so that it is possible 
to calculate 
mesoscale properties from monomeric-scale information. 
For the intramolecular structure factor $\omega_{\alpha}^{mm}(k)$, we adopt the 
Gaussian intrachain
distribution of Eq.(\ref{intra}).
The ``monomer"-c.o.m.\ intramolecular structure factor $\omega_{\alpha}^{mC}(k)$, 
can be  approximated well in {\it k}-space by a  Gaussian distribution\cite{yamakawa} as
\begin{eqnarray}
\omega_{\alpha}^{mC}(k)=N_{\alpha}e^{-k^2R_{g\alpha}^2/6} \ ,
\label{wmcGM}
\end{eqnarray}
with $\alpha \in \{A,B\}$. 
Including Eqs.(\ref{OZ},\ref{intra},\ref{wmcGM}) into Eqs.(\ref{hcc})
yields the following expressions for the total correlation functions between 
the c.o.m.\ of two chains
in the mixture
\begin{eqnarray}
h_{AA}(r)& = & \frac{1-\phi}{\phi}I^{\phi}_{AA}(r)+
\gamma^{2}I^{\rho}_{AA}(r) \ ,
\nonumber \\
h_{BB}(r) & = & \frac{\phi}{1-\phi}I^{\phi}_{BB}(r)+
\gamma^{-2}I^{\rho}_{BB}(r) \ ,
\label{h}\\
h_{AB}(r) & = & -I^{\phi}_{AB}(r)+
I^{\rho}_{AB}(r) \ ,
\nonumber
\end{eqnarray}
where $I^{\phi}_{\alpha\beta}(r)$ and $I^{\rho}_{\alpha\beta}(r)$ identify 
the concentration and density fluctuation contributions, respectively.
We introduce here a compact notation with the function 
$I^{\lambda}_{\alpha\beta}(r)$ defined as
\begin{eqnarray}
&I^{\lambda}_{\alpha\beta}(r)&= \nonumber \\
&&\frac{3}{4}\sqrt{\frac{3}{\pi}}\frac{\xi_{\rho}^\prime}
{R_{g\alpha\beta}}\vartheta_{\alpha\beta1}
\left(1-\frac{\xi_{c\alpha\beta}^2}{\xi_{\lambda}^2}\right) e^{
  -3r^2/(4R_{g\alpha\beta}^2)}
-\frac{1}{2}\frac{\xi_{\rho}^\prime}{r}\vartheta_{\alpha\beta2}
\left(1-\frac{\xi_{c\alpha\beta}^2}{\xi_{\lambda}^2}
  \right)^2 e^{
  R_{g\alpha\beta}^2/(3\xi_{\lambda}^2)} \nonumber \\
&&\times\left[ e^{r/\xi_{\lambda}}\mbox{erfc} 
\left(\frac{R_{g\alpha\beta}}{\xi_{\lambda} \sqrt{3}}+
 \frac{r\sqrt{3}}{2R_{g\alpha\beta}}\right)-
e^{ -r/\xi_{\lambda}}\mbox{erfc}\left(\frac{R_{g\alpha\beta}}{
\xi_{\lambda}\sqrt{3}}- \frac{r\sqrt{3}}{2R_{g\alpha\beta}} \right) \right]
\label{Ta}
\end{eqnarray}
and
\begin{eqnarray}
&\vartheta_{\alpha\beta1}&=
\frac{\left(1-\frac{\xi_{c\alpha\alpha}^2\xi_{c\beta\beta}^2}
{\xi_{c\alpha\beta}^2\xi_{\lambda}^2}\right)}
{\left(1-\frac{\xi_{c\alpha\beta}^2}{\xi_{\lambda}^2}\right)},\\
&\vartheta_{\alpha\beta2}&= 
\frac{\left(1-\frac{\xi_{c\alpha\alpha}^{2}}{\xi_{\lambda}^2}\right)
\left(1-\frac{\xi_{c\beta\beta}^2}{\xi_{\lambda}^2}\right)}
{\left(1-\frac{\xi_{c\alpha\beta}^2}{\xi_{\lambda}^2}\right)^2},
\label{mix}
\end{eqnarray}
where $\xi_{\lambda}\in \{\xi_{\phi},\xi_{\rho}\}$ and $\xi_{\rho}'=3/(\pi\rho\sigma_{AB}^2)$.
Radii of gyration in the blend are defined  as $R_{g \alpha \beta} = \sqrt{(R_{g\alpha}^2  
+ R_{g\beta}^2)/2}=\xi_{c\alpha \beta}\sqrt{2}$ so that, for example, if 
$\alpha=\beta=A$ then $R_{gAA} = R_{gA}=\xi_{cA}\sqrt{2}$ as in Eq.(\ref{xic}).

Eqs.(\ref{h}) can be further simplified 
in particular cases. For example, the {\it density fluctuation} 
contribution in both {\it self 
terms} is formally identical to the total distribution function
between the c.o.m.\ of two interacting polymers in a melt
\begin{eqnarray}
I^{\rho}_{AA}(r)& = &h(r)=\frac{3}{4}\sqrt{\frac{3}{\pi}}\frac{\xi_{\rho}^\prime}{R_g}
\left(1-\frac{\xi_c^2}{\xi_{\rho}^2}\right) e^{-3r^2/(4R_g^2)}
-\frac{1}{2}\frac{\xi_{\rho}^\prime}{r}
\left(1-\frac{\xi_c^2}{\xi_{\rho}^2}\right)^2
  e^{R_g^2/(3\xi_{\rho}^2)}
  \label{hccmelt} \\
&& \times\left[ e^{r/\xi_{\rho}}\mbox{erfc}
\left(\frac{R_g}{\xi_{\rho} \sqrt{3}}+\frac{r\sqrt{3}}{2R_g}\right)
-e^{ -r/\xi_{\rho}}\mbox{erfc}\left(\frac{R_g}{
\xi_{\rho}\sqrt{3}}- \frac{r\sqrt{3}}{2R_g} \right) \right] .
\nonumber
\end{eqnarray}
This expression can by reduced to a more compact form in the limit of large 
molecules ($N\rightarrow\infty$) by
expanding the total correlation function as a function of the  vanishing parameter
$\xi_{\rho}/R_g\rightarrow0$
\begin{eqnarray}
h(r)\approx  -
\frac{39\sqrt{3}}{16\sqrt{\pi}}\frac{\xi_\rho}{R_g}
\left(1+\frac{\xi_\rho}{\xi_c}\right)\left[1-\frac{9r^2}{26R_g^2}\right]
e^{-3r^2/(4R_g^2)} = I^{\rho}(r,\xi_{\rho}/R_g\rightarrow0) \ .
\label{Troappr}
\end{eqnarray}
Since our initial equations of intra- and intermolecular structure 
factors rely on the assumption of
Gaussian-chain statistics, a condition which formally 
holds only for polymer chains of infinite length,
Eq.(\ref{Troappr}) is entirely consistent with the general 
description of our system and can be
adopted in general for $N \ge 30$.

The density fluctuation contribution due to 
{\it cross interactions} $AB$ cannot be simplified
to the melt expression, unless the blend is totally symmetric 
($\sigma_A=\sigma_B$ and $N_A=N_B$) or only slightly asymmetric. 
In the totally symmetric case,
$I^{\rho}_{AA}=I^{\rho}_{BB}=I^{\rho}_{AB}=h(r)$.
If the system is only slightly asymmetric, the condition 
$\vartheta_{AB1}\approx \vartheta_{AB2}$
is still fulfilled, but only  
for large chains ($N\ge 30$) and  $\mu=0.1-2$ while $1\le\gamma\le2$.
For this case, Eqs.(\ref{h}) simplify to
\begin{eqnarray}
I^{\rho}_{\alpha\beta}(r) \approx
\vartheta_{\alpha\beta1}I^{\rho}(r,\xi_{\rho}/R_{g\alpha\beta}\to0)
\label{hAB1} \ ,
\end{eqnarray}
with $I^{\rho}(r,\xi_{\rho}/R_{g\alpha\beta}\to0)$ defined by Eq.(\ref{Troappr}).

For the {\it concentration fluctuation} term,
the problem is further complicated by the fact that the functions
in Eqs.(\ref{h})
depend not only on the parameter $\xi_{\rho}/R_g \rightarrow 0$, but also on
the additional parameter
$\xi_{\phi}/R_g$ which
does not approach zero either
in the athermal ($\xi_{\phi}\rightarrow \xi_c\approx R_g$) or in the spinodal 
($\xi_{\phi}\rightarrow \infty$)
regions.
However, at the spinodal, where the concentration fluctuation lengthscale diverges,
the expansion of 
$I^{\phi}_{\alpha\beta}(r)$ 
in the limit of $R_{g\alpha\beta}/\xi_{\phi}\to0$ yields:
\begin{eqnarray}
I^{\phi}_{\alpha\beta}(r)&\approx&
\frac{\xi_{\rho}^{\prime}}{R_{g\alpha\beta}}\left[\frac{3}{4}\sqrt{\frac{3}{\pi}}
e^{-3r^{2}/(4R_{g\alpha\beta}^{2})}+\frac{R_{g\alpha\beta}}{r}
\mbox{erf}\left(\frac{\sqrt{3}}{2}\frac{r}{R_{g\alpha\beta}}\right)
-\frac{R_{g\alpha\beta}}{\xi_{\phi}}+{\mathcal O}\left(\frac{R_{g\alpha\beta}^2}
{\xi_{\phi}^2}\right)\right]
\label{sconc} \ ,
\end{eqnarray}
where we used the fact that $\lim_{\xi_{\phi}\to\infty}{\vartheta_{\alpha\beta i}=1}$
with $i \in \{1,2\}$.
The first two terms on the r.h.s. of Eq.(\ref{sconc}) describe the spinodal 
decomposition and correspond
to the maximum possible contribution to the pair distribution functions due to concentration
fluctuations.
When the system approaches its spinodal decomposition, 
contributions in $R_{g\alpha\beta}/\xi_{\phi}$
become increasingly small in magnitude, to the point  of
no longer compensating the
first two contributions on the r.h.s.\ of the equation. 
In the self terms of the total correlation
function, this phenomenon manifests itself as an overshoot
in the function, which corresponds to clustering
of like species, or attractive interactions between molecules of the same type.
Meanwhile, the cross correlation decreases, showing the beginning of phase 
separation (refer to model calculations in Sec.\ VI).

In the {\it athermal} limit where $\chi/\chi_s \rightarrow 0$, 
if the blend is asymmetric,
entropic contributions can induce demixing for the mesoscale
liquid structure both through density and concentration fluctuation contributions.
However, if the blend is 
symmetric ($\sigma_A=\sigma_B$ and $N_A=N_B$), only density fluctuation
contributions ($I^{\rho}$) govern the properties of the mixture, and 
$I^{\phi}_{AA}=I^{\phi}_{BB}=I^{\phi}_{AB}=0$.
For symmetric blends,
the total correlation functions recover
the expression for the melt correlation function.\cite{meltprl}

\section{Chain-Overlap and Intermolecular Mean-Force Potential}
The overlap value of intermolecular total pair distribution functions provides information
on the degree of interpenetration between two chains as a function of monomer density, 
temperature, blend composition, chain degree of polymerization, 
and asymmetry ratios ($\gamma$ and $\mu$).
In general, the intermolecular potential of mean force for two particles in a liquid  
is given by $W_{\alpha\beta}(r)=-k_BT \ln{g_{\alpha\beta}(r)} 
\approx -k_BT h_{\alpha\beta}(r)$.
For polymer melts, the (repulsive) potential at contact $W(0)$ is known to decrease  
with increasing  
degree of polymerization, increasing density, or increasing 
polymer stiffness.\cite{Hansen,marinamacrom}
For polymer blends, the situation is more complex since the mismatch in local structure and 
chain length between the two components can give relevant concentration fluctuation 
contributions
even in the athermal regime. 

At contact, the blend {\it density fluctuation} term is similar to the melt case. In the limit 
$r\to0$ and  $N\to\infty$, Eq.(\ref{hccmelt}) reduces to
\begin{eqnarray}
\lim_{r\to0}I^{\rho}_{\alpha\beta}(r)&\approx&  
-\frac{39}{16}\sqrt{\frac{3}{\pi}}\frac{3}{\pi\rho\sigma_{AB}^2 R_{g\alpha\beta}} =
I^{\rho}(0,\xi_{\rho}/R_{g\alpha\beta}\to0)
\label{Tro0}
\end{eqnarray}
with $\vartheta_{\alpha\beta i}=1$ and $i \in \{1,2\}$. 
Eq.(\ref{Tro0}) is valid for self correlation functions, and for the cross correlation
functions if the system is symmetric or slightly asymmetric.
The {\it concentration fluctuation} term in the limit $r\to0$ also simplifies to
\begin{eqnarray}
&&\lim_{r\to0}{I^{\phi}_{\alpha\beta}(r)} \approx
\frac{3}{4}\sqrt{\frac{3}{\pi}}\vartheta_{\alpha\beta 1}\frac{3}
{\pi\rho\sigma_{AB}^2R_{g\alpha\beta}}
\left(1-\frac{\xi_{c\alpha\beta}^2}{\xi_{\phi}^2}\right)+ \nonumber\\
&&+\vartheta_{\alpha\beta2}\frac{3}{\pi\rho\sigma_{AB}^2R_{g\alpha\beta}}
\left(1-\frac{\xi_{c\alpha\beta}^{2}}{\xi_{\phi}^{2}}\right)^{2}\left[\sqrt{\frac{3}{\pi}}-
\frac{R_{g\alpha\beta}}{\xi_{\phi}}e^{R_{g\alpha\beta}^{2}/(3\xi_{\phi}^{2})}
\mbox{erfc}\left(\frac{1}{\sqrt{3}}\frac{R_{g\alpha\beta}}{\xi_{\phi}}\right)\right]
\ .
\label{Tfi0}
\end{eqnarray}
Summarizing, the total correlation functions at contact for structurally  
symmetric ($\vartheta_{AB1}=1$)
and slightly asymmetric ($\vartheta_{AB1}\approx 1$)
polymer blends are written in short-hand notation as 
\begin{eqnarray}
\lim_{r\to0} h_{AA}(r)\approx\frac{1-\phi}{\phi}I^{\phi}_{AA}(0)+
\gamma^{2}I^{\rho}(0,\xi_{\rho}/R_{gAA}\to0) \ ,
\nonumber \\
\lim_{r\to0} h_{BB}(r)\approx\frac{\phi}{1-\phi}I^{\phi}_{BB}(0)+
\gamma^{-2}I^{\rho}(0,\xi_{\rho}/R_{gBB}\to0)  \ ,
\label{h0}\\
\lim_{r\to0} h_{AB}(r)\approx-I^{\phi}_{AB}(0)+
\vartheta_{AB1}I^{\rho}(0,\xi_{\rho}/R_{gAB}\to0) \ .
\nonumber
\end{eqnarray}
For both density and concentration fluctuation contributions, each total 
distribution function
at contact
depends on density, stiffness, and degree of polymerization as $I^{\rho,\phi}\propto
1/(\rho\sigma_{AB}^2R_{g\alpha\beta})$, essentially recovering the melt behavior:
an increase in polymer-polymer contact is observed  with increasing 
degree of polymerization ($R_g$),
stiffness ($\sigma$ and $R_g$), and liquid density ($\rho$).

\section{Mapping blend correlation functions onto a mesoscopic soft-colloidal description}
Our analytical c.o.m.\ correlation functions effectively map a polymer blend onto a fluid of 
interacting soft-colloidal particles. To test the quality of our renormalized description
we compare its predictions  with known properties of colloidal mixtures.\cite{HansenMcd}
 
At the mesoscale, a polymer blend is a two-component mixture of colloidal
particles of type $A$ and $B$ with volume fraction $\phi=n_A/(n_A + n_B)$, total density 
$\rho_{ch}=\rho/N$, and
particle size $R_{gA}$ and $R_{gB}$. When species $A$ 
is chosen to be our reference system, the size mismatch parameter is
$\gamma=R_{gB}/R_{gA}$, and the reduced chain density is $\rho_{ch}^*=\rho_{ch}R_{gA}^3$.
This renormalized description can be formally obtained from Eqs.(\ref{hcc}) by setting
the effective number 
of statistical segments in the two components
to be equal, $N_A=N_B=N$, while the chain asymmetry is completely accounted for by
the different statistical
segment lengths
$\sigma'_A=R_{gA}\sqrt{6/N}$ and $\sigma'_B=R_{gB}\sqrt{6/N}$.
Each effective segment
includes the effect of branching and chain semiflexibility, and must be
equal or larger than its
polymer persistence length, a condition fulfilled by long semiflexible chains.

It is convenient to define universal renormalized quantities 
which simplify the final equations.
The normalized density fluctuation correlation length is
$\tilde{\xi}_{\rho} =\xi_{\rho}/R_{gA}$, the normalized
concentration fluctuation correlation length is
$\tilde{\xi_\phi}=  \xi_\phi/R_{gA}=\{[1-\phi(1-\gamma^2)]/[2(1- \chi/\chi_s)]\}^{1/2}$,
and $\tilde{\xi}_{\rho}' =\xi_{\rho}'/R_{gA}=
\left\{2 \pi \rho_{ch}^* [\phi+(1-\phi)\gamma^2)]\right\}^{-1}$.
The total pair distribution functions Eqs.(\ref{Ta}) now depend
only on mesoscopic reduced variables, including  the space coordinate $\tilde{r} = r/R_{gA}$
as
\begin{eqnarray}
I^{\lambda}_{\alpha \beta} (\tilde{r})&=&
\frac{3}{4}\sqrt{\frac{3}{\pi}}\frac{\tilde{\xi}_{\rho}'}{a} 
\left(1-\frac{a^2}{2\tilde{\xi}_{\lambda}^2}\right) b_1
e^{-3\tilde{r}^2/(4a^2)} -\frac{1}{2}\frac{\tilde{\xi}_{\rho}'}{\tilde{r}}
\left(1-\frac{a^2}{2\tilde{\xi}_{\lambda}^2}\right)^2 b_2
e^{a^2/(3\tilde{\xi}_{\lambda}^2)} \nonumber \\
&&\times\left[e^{\tilde{r}/\tilde{\xi}_{\lambda}}\mbox{erfc}
\left(\frac{a}{\tilde{\xi}_{\lambda}\sqrt{3}
}+\frac{
\tilde{r}\sqrt{3}}{2 a}\right)-
e^{-\tilde{r}/\tilde{\xi}_{\lambda}}\mbox{erfc}
\left(\frac{a}{\tilde{\xi}_{\lambda}\sqrt{3}}-
\frac{\tilde{r}\sqrt{3}}
{2 a}\right)\right] \ ,
\end{eqnarray}
where $\tilde{\xi}_{\lambda} \in \{\tilde{\xi}_{\rho}, \tilde{\xi}_{\phi}\}$.
Here $a=1$ if  $\alpha=\beta=A$,
$a=\gamma$ if $\alpha=\beta=B$, 
and $a=\sqrt{(1+\gamma^2)/2}$ for the cross terms.
Also $b_1=b_2=1$ for the self terms, while for the cross terms
\begin{eqnarray}
b_1=\frac{\tilde{\xi}_{\lambda}^2 -\gamma^2/(1+\gamma^2)}
{\tilde{\xi}_{\lambda}^2 -(1+\gamma^2)/4} \ ,
\end{eqnarray}
and
\begin{eqnarray}
b_2=\frac{(2 \tilde{\xi}_{\lambda}^2 -1)(2 \tilde{\xi}_{\lambda}^2 -\gamma^2)}
{[ 2 \tilde{\xi}_{\lambda}^2 -(1+\gamma^2)/2]^2} \ .
\end{eqnarray} 
To make contact with the theory of colloidal particle mixtures (e.g., liquid alloys) it
is convenient to analyze the properties of the renormalized polymer blend in reciprocal
space. The static structure factors for the mixture are defined as
\begin{eqnarray}
S_{AA}(k) & = & \phi + \phi^2 \rho_{ch} h_{AA}(k) \nonumber \ , \\
S_{BB}(k) & = & 1-\phi + (1-\phi)^2 \rho_{ch} h_{BB}(k) \ , \\
S_{AB}(k) & = & \phi(1-\phi) \rho_{ch} h_{AB}(k) \nonumber \ .
\end{eqnarray}
Linear combinations of these functions describe fluctuations in 
number density and concentration,
following Bhatia and Thornton's formalism.\cite{Bhatia}
The density fluctuation contribution ($S^{NN}$ in the conventional notation for
``metal alloys") is given by
\begin{eqnarray}
S^{\rho \rho}(k)=S_{AA}(k)+S_{BB}(k)+2S_{AB}(k) \ , \label{sroro}
\end{eqnarray}
while the concentration fluctuation contribution ($S^{CC}$) is
\begin{eqnarray}
S^{\phi \phi}(k) =(1-\phi)^2 S_{AA}(k)+ \phi^2 S_{BB}(k)- 
2\phi (1-\phi) S_{AB}(k) \ ,  \label{sfifi}
\end{eqnarray}
and their coupling ($S^{CN}$) is 
\begin{eqnarray}
S^{\rho \phi}(k)=(1-\phi) S_{AA}(k)- \phi S_{BB}(k)+
(1-2 \phi) S_{AB}(k) \ . \label{srofi}
\end{eqnarray}
Analytical formulas for Eqs.(\ref{sroro}-\ref{srofi}) are readily 
obtained using Eqs.(\ref{hcc}).
Their model calculations are presented in Fig.2. 
Eqs.(\ref{sroro}-\ref{srofi}) follow closely the
behavior in reciprocal space observed for colloidal mixtures.
For example, $S^{\rho \rho}(k)$ has a $k$-dependence similar to the static structure factor
for a single-component liquid. However,  
since our colloids are soft and can interpenetrate, 
there is no formation of solvation shells in the
mixture and $S^{\rho \rho}(k)$ becomes a monotonically-increasing function of $k$.
$S^{\phi \phi}(k)$ and $S^{\rho \phi}(k)$ oscillate about the values $\phi (1-\phi)$ and zero,
respectively, as observed in colloidal mixtures with oscillations becoming less pronounced
in $S^{\phi \phi}(k)$.
A more intuitive picture of the density-concentration fluctuation coupling term can be 
established by rewriting it as\cite{Bhatia}
\begin{eqnarray}
S^{\phi \rho}(k) = \phi (1-\phi) \rho_{ch} 
\int [P_{A}(r)-P_{B}(r)] \frac{\mbox{sin}(kr)}{kr} 4 \pi r^2 dr
\end{eqnarray}
where $P_{\alpha}(r)=(1-\phi)g_{B\alpha}(r) + \phi g_{A\alpha}(r)$ is the probability of 
encoutering particle clustering
of species $A$ or $B$ around the colloid $\alpha \in \{A$,$B\}$. 
In this way, the function $S^{\phi \rho}(k)$ represents
a measure of the difference in local clustering between 
species $A$ and $B$. Maxima and minima in the
function point at lengthscales characterized by large asymmetry in the liquid structure 
due to the mismatch in 
particle size. If the two species are identical,
the mixture is uniform and $S^{\phi \rho}(k)=0$ for any $k$.

In the $k \rightarrow 0$ limit, the density fluctuation contribution and its coupling with
concentration fluctuations reduce to the
simplified forms
\begin{eqnarray}
S^{\rho \rho}(0)& = & \frac{\xi_{\rho}^2}{\xi_{cA}^2} 
\frac{\phi \gamma^2 + 1 -\phi}{\gamma^2} \  , 
\label{srr}\\
S^{\rho \phi}(0)& = & \phi (1-\phi) \frac{\gamma^2 -1}{\gamma^2}
\frac{\xi_{\rho}^2}{\xi_{cA}^2} \ , \label{srf}
\end{eqnarray}
while the concentration fluctuation contribution is
\begin{eqnarray}
S^{\phi \phi}(0)& = & \frac{\phi (1-\phi)}{1-\chi/\chi_s} + 
\frac{\phi^2 (1-\phi)^2(\gamma^2 -1)^2}
{(\phi \gamma^2 + 1 -\phi)\gamma^2}\frac{\xi_{\rho}^2}{\xi_{cA}^2} \ .  \label{sff}
\end{eqnarray}
For a blend of symmetric polymers, where $R_{gA}=R_{gB}$ and $\gamma=1$,
our equations become completely consistent with the theory for a mixture of symmetric
colloidal particles:\cite{Bhatia,HansenMcd}
Eq.(\ref{srr}) simplifies to $S^{\rho \rho}(0) = (\xi_{\rho}/\xi_{cA})^2$ 
the melt compressibility,
$S^{\rho \phi}(0)=0$, and the concentration fluctuation contribution
becomes 
\begin{eqnarray}
S^{\phi \phi}(0)
& = & \frac{\phi (1-\phi)}{1-2 \phi (1-\phi)\Delta E} \ .
\end{eqnarray}
Here we introduced Flory's definition of the spinodal $\chi_s$, and the
renormalized $\chi$ parameter for the coarse-grained polymer mixture
$\Delta E= N\chi = N\epsilon_{AB} - (N \epsilon_{AA}+ N \epsilon_{BB})/2$.
For long polymer chains or high density, Eqs.(\ref{srr},\ref{srf}) vanish 
since $\xi_\rho/\xi_c \rightarrow 0$, 
while the concentration fluctuation contribution yields
$S^{\phi \phi} (0)=\phi (1-\phi)/(1-\chi/\chi_s)$.

In general, asymmetry between the two colloidal species is quantified by the 
dilation factor\cite{KirkwoodBuff}
\begin{eqnarray}
\delta & = & \frac{v_A-v_B}{\phi v_A+(1-\phi) v_B}= 
\frac{S^{\rho \phi}(0)}{S^{\phi \phi}(0)} \\
& = & \frac{(\phi \gamma^2 + 1 -\phi)(\gamma^2 -1)}{\phi (1-\phi) 
(\gamma^2 -1 )^2+ \gamma^2 
\xi_{\phi}^2/\xi_{\rho}^2} \ . \nonumber
\end{eqnarray}
If the partial molar volumes per particle 
$v_{\alpha}=(\partial V/\partial n_\alpha)_{n_{\beta \neq \alpha},P,T}$ 
of the two species are
identical ($\delta =0$ and $\gamma=1$) there is no correlation 
between the fluctuations in particle
number and concentration and
$S^{\rho \phi}(k)=0$. At the spionodal, $S^{\phi \phi}(k)$ diverges 
and $\delta \rightarrow 0$.

The
isothermal compressibility for a colloidal mixture is defined\cite{KirkwoodBuff} as
\begin{eqnarray}
\rho_{ch} k_B T \kappa_T & = & S^{\rho \rho}(0)-\frac{S^{\rho \phi}(0)^2}{S^{\phi \phi}(0)}
= S^{\rho \rho}(0) - \delta^2  S^{\phi \phi}(0) \ .
\end{eqnarray}
Here,
\begin{eqnarray}
\rho_{ch} k_B T \kappa_T & = & \frac{\xi_{\rho}^2}
{\xi_{cA}^2} \frac{\phi \gamma^2 + 1 -\phi}{\gamma^2}
\left[1-\frac{(\gamma^2 -1)^2 \phi (1-\phi)}{\phi (1-\phi)
(\gamma^2 -1)^2/\gamma^2+\xi_\phi^2/\xi_\rho^2}\right] \ , 
\end{eqnarray}
recovers the melt compressibility when  the system is composed 
of colloidal particles of identical size,
$\gamma=1$.
The compressibility is slightly temperature-dependent through the correction contribution
due to $S^{\rho \phi}(0)^2/S^{\phi \phi}(0)$. The latter, however, is small for large
polymer chains since it scales with degree of polymerization
as $N^{-1}$ in the athermal regime, and vanishes approaching the spinodal curve where 
the concentration correlation
length diverges. This is true at any lengthscale and the $k$-dependent blend 
compressibility can be approximated
by its first contribution $S^{\rho \rho}(k)$ for the entire range of $k$ and for blends of
polymer chains with symmetric or
asymmetric size, in agreement with colloidal particle mixtures (Fig.2).

\section{Center-of-Mass Distribution Functions: Model Calculations and Comparison with UA-MD Computer Simulations}
In this section
we present some predictions of our theoretical approach
for the clustering and interpenetration of polymer chains in a blend as a function of
chain length, semiflexibility, polymer volume fraction, and $\chi$ dependence
by model calculations.
In our model calculations, unless otherwise specified, $N_A=500$, $\rho_m=0.03$ \AA$^{-3}$,
and $\sigma_A=3.0$ \AA. These parameters have been chosen to be 
consistent with the UA-MD simulations
of
polyolefin blends presented below. The chain length,
however, has been chosen to be quite large to avoid finite-size effects which could
veil the general trend of blend properties.
Calculations performed ``close to the spinodal condition" are for a
$\chi$ parameter which deviates $0.01 \% $ from $\chi_s$.

As a first study, we look at a structurally symmetric blend for which
$\gamma=1$ and $N_A=N_B$, while
varying the blend composition $\phi\in\{0.25,0.50,0.75\}$.
In the \textit{athermal} limit
$\chi \ll \chi_s$,
all distribution functions become identical for any blend composition $\phi$.
However, when concentration fluctuations develop (Fig.3), we observe
an enhanced clustering and increased number of self contacts in the \textit{minority}
(low $\phi$) species, which
corresponds to a strong
attractive interaction in the mean-force potential.
The \textit{majority} (high $\phi$) species
increases the number of self-contacts, but still resembles in shape
its athermal limit. This effect corresponds to the initial stage of droplet 
formation in the demixing of
compositionally asymmetric mixtures.
In general, the extent
of demixing 
increases with increasing asymmetry in blend composition.

To investigate the effect of mismatch in chain flexibility, we analyze blends of
polymers with identical degree of polymerization ($N_A=N_B$) where component $A$ is
flexible while the stiffness of component $B$ is increasing (increasing $R_{gB}$ and
$\gamma\in\{1.0,1.2,1.5\}$).
The data from UA-MD simulations analyzed subsequently in this paper (see Table)
belong to this group.
In general, we observe that an increase
in $\gamma$ enhances the number of $BB$ and $AB$ contacts and reduces the number of 
$AA$ contacts (Fig.4).
This effect is observed in both athermal and
thermal mixtures.
A stiff chain has a larger $R_g$ than a flexible
molecule with the same degree of polymerization. In this case, the 
number of chains that can occupy
the volume spanned by the stiff component increases with increasing $R_g$.
Furthermore, stiff chains pack better at short distances due to geometric effects
(stretched configurations enhance
$BB$ contacts), while flexible chains, which have a coiled configuration,
tend to pack more efficiently at distances of 
order $R_{gA}$. Overall, these two effects lead to an enhanced miscibility in
mixtures of polymers with different flexibility and same degree of polymerization.

The potential of mean-force for chains of equal length is found to decrease with increasing
degree of polymerization
in agreement with
computer simulations.\cite{Hall,Hansen}
The effect of mismatch in chain length for polymer species having the same semiflexibility
is studied using a blend of polymers  with
$N_{A}=500$ fixed and  $N_{B}\in\{250,500,1000\}$, corresponding to $\mu\in\{0.5,1.0,2.0\}$
(Fig.5).
Analogous effects are observed in both the athermal and the thermal limits. In the athermal
limit, the mismatch in chain size prevents the random mixing of the two 
components and ``entropic"
effects appear.
Trivially, as the $R_g$ of the $B$ component increases the number of chains
that can interpenetrate
in the volume spanned by $B$ increases. This leads to a larger number of
intermolecular contacts for all pair distribution functions
$AA$, $BB$, and $AB$. This simple trend is combined with a more subtle behavior:
a mismatch in chain length favors interpenetration between different species, 
so that the number of
$AB$ contacts is always higher than the number of $AA$ and $BB$ contacts except for chains
of the same length. This property becomes more pronounced when the difference in chain length
increases.
Also, short flexible chains tend to pack efficiently on the scale of 
their $R_g$, while these effects
are averaged out in large flexible chains, where ``solvation shells" 
do not appear (a
behavior analogous to the monomer distribution functions).
Our theoretical predictions are in qualitative agreement with off-lattice 
Monte Carlo simulations
of polymer solutions.\cite{Hall}

To investigate in more detail these effects, we analyze polymer blends
which have  \textit{constant} $R_{gA}$ and $R_{gB}$. The
change in chain length in the $B$ component is  correlated with a
change in semiflexibility. For example, the $A$ component is
mixed with the component $B$, which   is both
shorter and stiffer than $A$ ($\mu=2.00$ and $\gamma=1.88$), or
with a chain $B$  larger than $A$ but with comparable
flexibility ($\mu=0.66$ and $\gamma=1.09$). 
In analogy with our previous observation, we see (Fig.6) that
stiff chains tend to pack more efficiently than flexible ones at short distances, increasing
in this way the number of $BB$ and $AB$ contacts. As a consequence, 
the number of $AA$ contacts
in a mixture of stiff $B$ chains is decreased with respect to the mixture of flexible
polymers.  Close to the spinodal, flexible chains ($A$ and $B$) are in 
coiled configurations which
tend to pack more efficiently at the distance of their $R_g$. Stiff chains, instead, tend to
interpenetrate more efficiently and the number of $BB$ contacts strongly  
increases in the mixture
containing short and stiff $B$ chains.

Next, we compare our
analytical expressions against data from UA-MD simulations of polymer 
blends described in the Table.\cite{Grest,grest} For each blend considered,
the theory agrees with simulation data fairly well. However,
we present here only few systems
which are representative of our calculations.
Input parameters to the theory and the procedure used in the calculations
were discussed in Section II.

The c.o.m.\ total correlation function provides an estimate of the number of
molecules interpenetrating at some relative distance $r$. In all the plots (Fig.7)
we observe that chains of the stiffest component ($B$), which have 
extended configurations, tend to pack at short distances $r < R_{gB}$ more efficiently
than flexible ones. 
Flexible molecules, which have coiled
configurations, show  a high (low) number of ``intramolecular" (intermolecular) 
contacts and pack most efficiently 
at distances on the order of the
overall polymer size,
$r \approx 1.5 R_{gA}$.
The extent of intermolecular chain packing upon blending also depends on 
polymer flexibility. The theory predicts that
the stiff (flexible)
component packs better (worse),
and the number of self contacts increases (decreases) when mixed with
a more flexible (stiffer) polymer, in agreement with simulations.\cite{Maranas}

In general, the agreement between theory and simulations is good with the
exception of the PIB/hhPP and PIB/iPP blends for which agreement is only
qualitative. For these systems the theory overestimates the 
number of intermolecular
contacts. However, it is well known that PIB blends
are usually immiscible blends at
these temperatures and for these chain lengths.\cite{grest,Krishnamoorti} 
PIB exhibits very efficient intra- and intermolecular packing which leads
to a thermal expansion
coefficient and an isothermal compressibility smaller than in other
polyolefin  blends.\cite{Krishnamoorti}
This behavior is due to strong 
attractive interactions between the methyl ($-\mbox{CH}_3$) groups, which in PIB molecules 
are in very large number 
(about 50\% of the total number of united atoms).\cite{Freed2}
However, even for PIB blends, the theory shows very good agreement with simulation data
for $r \approx R_g$.

Finally, in Fig.8 we show how the theory can predict the correct qualitative 
behavior for blends following a LCST phase diagram. At high temperature the blend
demixes and shows a distribution function for the $AB$ component that is consistently
lower than the $AA$ and the $BB$ contributions. At low temperature, instead, the 
mixing of the $A$ and $B$ species is enhanced with the $AB$ function being always higher than
its $AA$ and $BB$ counterparts.

\section{Conclusions}
In this paper we present the first analytical solution for the center-of-mass
total distribution functions
for polymer mixtures. We start from first-principles liquid state theory
and derive
a set of equations which describe c.o.m. distribution functions between
pairs of interacting molecules in the mixture. The mesoscale description of
the liquid so obtained
depends on local chemical parameters, such as the 
radius of gyration and bond length.
Moreover, mesoscale distribution functions depend on
thermodynamic parameters such as density, blend composition, and proximity of the
system to its spinodal decomposition.

The analytical c.o.m. total distribution functions
quantitatively reproduce data from microscopic
united-atom molecular
dynamics simulations without the need of fitting parameters.
In this comparison, input to the theory are data from the UA-MD simulations.
The excellent agreement between analytical theory and simulations in both
real and reciprocal space is an indication that our approach provides
a reliable mesoscale description of polymer mixtures.

Moreover, our formalism recovers the known analytical functions for
the fluctuations in number density and
concentration (structure factors) for interacting soft-colloidal mixtures.
The one-to-one agreement between the two formalisms supports
the validity of our approach as a procedure to map polymer blends onto systems
of interacting soft-colloidal particles.

\section{Acknowledgments}
We are grateful to G.\ S.\ Grest and E. Jaramillo for sharing UA-MD computer simulation
trajectories. We acknowledge support
from the National Science Foundation
under grant DMR-0207949. Also, EJS acknowledges support from an NSF Graduate
Research Fellowship.

\pagebreak

\pagebreak
{\bf Table:} Polyolefin blends ($T=453$ K, $N_A=N_B=96$).\\

\begin{tabular}{|c|c|c|c|c|c|} \hline \hline

Blend[$A/B$] & $\phi$ & $\rho$ [sites/\AA$^3$] & $R_{gA}$ [\AA] & $\gamma$ & $\chi$\\
\hline
iPP/hhPP     & 0.50   & 0.0332                 & 11.28          & 1.09     
& $-0.00364+1.84/T$\cite{Balsara}\\
hhPP/sPP     & 0.50   & 0.0332                 & 12.18          & 1.14     
& $-$\\
PIB/iPP      & 0.50   & 0.0343                 &  9.68          & 1.16     
& $0.017+5.6/T$\cite{grest}\\
sPP/PE       & 0.50   & 0.0328                 & 13.89          & 1.19     
& $-$\\
iPP/sPP      & 0.50   & 0.0328                 & 11.37          & 1.22     
& $-$\\
PIB/hhPP     & 0.50   & 0.0343                 &  9.69          & 1.28     
& $0.027-11.4/T$\cite{grest}, see also\cite{Krishnamoorti,Balsara,Lee}\\
hhPP/PE      & 0.50   & 0.0332                 & 12.32          & 1.34     
& $-0.0294+17.58/T$\cite{Jeon}, see also\cite{Lee}\\
PIB/sPP      & 0.50   & 0.0343                 &  9.76          & 1.41     
& $-$\\
PIB/PE       & 0.50   & 0.0343                 &  9.76          & 1.68     
& $0.00257+4.99/T$\cite{Lee}\\
iPP/PE       & 0.25   & 0.0328                 & 11.35          & 1.47     
& $0.005$\cite{Grest}\\
iPP/PE       & 0.75   & 0.0328                 & 11.33          & 1.48     
& $0.01$\cite{Grest}\\
\hline\hline
\end{tabular}
\pagebreak

FIG.\ 1: Plot of $n_{\alpha\beta}(r)$ against $r$ for the hhPP/PE blend.
Theoretical curves for athermal (full lines) and thermal (dashed lines)
conditions are compared with simulation data (symbols): $AA-$ (circles), 
$AB-$ (diamonds, inset), and
$BB-$terms (squares). 
Arrow indicates the lengthscale $R_{gA}$.

FIG.\ 2: Plot of colloid partial structure factors against $kR_{gA}$ for
$\gamma\in$ \{1.0 (dashed lines), 1.5 (full lines)\} 
with $\mu=1.0$ and $\phi=0.5$. 
Left panel. Upper portion: 
$S^{\phi\phi}(k)$ with $\chi/\chi_s\in$ \{0.0, 0.2, 0.4, 0.6, 0.8\}
from bottom to top. Lower portion: $S^{\rho\phi}(k)$ with $\chi/\chi_s=
\mbox{0.0}$ (top line) and $\chi/\chi_s=\mbox{0.8}$ (bottom line). 
Right panel: $S^{\rho\rho}(k)$ for $\chi/\chi_s\in$ \{0.0, 0.2, 0.4, 0.6, 0.8\}
(curves are
indistinguishable in the plot). Right panel also shows 
$S^{\rho\rho}(k)-\delta^2S^{\phi\phi}(k)$ 
(dot-dashed lines) for $\gamma=1.5$.

FIG.\ 3: Plots of $h_{\alpha\beta}(r)$ against $r/R_{gA}$ for a structurally
symmetric blend ($\mu=1.0$ and $\gamma=1.0$) close to (0.01\% from) its
spinodal decomposition. 
Left panel: effect of blend composition for 
$AA-$term  at $\phi\in$ \{0.25 (long-dashed line), 0.50 (dot-dashed line),
0.75 (full line)\}. The behavior for $BB-$term is complementary to the
$AA-$term at $1-\phi$. Right panel: $AB-$term for which curves 
superimpose onto the
full line.
Also shown for comparison is the athermal case at
$\phi=0.5$ (dashed lines).

FIG.\ 4: Plots of $h_{\alpha\beta}(r)$ against $r/R_{gA}$ for $\gamma\in$
\{1.0 (dashed lines),
1.2 (full lines), 1.5 (dot-dashed lines)\} with $\mu=1.0$ and $\phi=0.5$
for athermal conditions. 
Left panel: $AA-$ (thin lines) and $BB-$terms
(thick lines). Right panel: $AB-$terms. Note that $AA=BB$ for the $\gamma=1$ case.

FIG.\ 5: Plots of $h_{\alpha\beta}(r)$ against $r/R_{gA}$ for $\mu\in$
\{0.5 (dot-dashed lines), 1.0 (dashed lines) 2.0 (full lines)\} with
$\gamma=1.0$ and $\phi=0.5$ under athermal conditions. 
Left panel: $AA-$terms; middle
panel: $AB-$terms; right panel: $BB-$terms. 

FIG.\ 6: Plots of $h_{\alpha\beta}(r)$ against $r/R_{gA}$ for systems with
different $\mu$ and $\gamma$, but constant radii of gyration $R_{gA}$ and
$R_{gB}$ at $\phi=0.5$ for athermal conditions: full lines for $\mu=2.00$
and $\gamma=1.88$; dashed lines for $\mu=1.00$ and $\gamma=1.33$; dot-dashed
lines for $\mu=0.66$ and $\gamma=1.09$. Left panel: $AA-$terms, middle panel:
$AB-$terms; right panel: $BB-$terms.

FIG.\ 7: Plots of $h_{\alpha\beta}(r)$ against $r/R_{gA}$ for blends.
Theoretical predictions in athermal (full lines) and thermal (dashed
lines) conditions are compared with UA-MD simulations of blends (symbols): $AA-$
(circles), $AB-$ (diamonds), and $BB-$terms (squares).  Blend parameters
are given in the Table.

FIG.\ 8: Plots of $h_{\alpha\beta}(r)$ against $r/R_{gA}$ for a LCST blend
described by $\chi=0.01125-4.75/T$ ($N_{A}=96$, $\rho_m=0.034$ \AA$^{-3}$, 
$\sigma_{A}=2.44$ \AA, $\phi=0.50$, and $\chi_s=0.021$).  Shown are $AA-$
(dot-dashed lines), $AB-$ (full lines), and $BB-$terms (dashed lines) at
two temperatures: $T=150$ K (thin lines) and $T=10,000$ K
(thick lines).

\end{document}